\documentclass[prb,floatfix,twocolumn,showpacs,amsmath,amssymb,
superscriptaddress]{revtex4}
\usepackage{graphicx}
\usepackage{dcolumn}
\usepackage{bm}
\begin{document}

\title{ Exact Jastrow-Slater wave function for the one-dimensional Luttinger model}
\author{Benjamin Tayo and Sandro Sorella}
\affiliation{Physics Department, Lehigh University - 16 Memorial E
Dr., Bethlehem, PA 18015, USA} \affiliation{INFM-Democritos,
National Simulation Center and International School for Advanced
Studies (SISSA), I-34014 Trieste, Italy}
\date{\today}

\begin{abstract}
We show that it is possible to describe the ground state of the
Luttinger model in terms of a Jastrow-Slater wave function.
Moreover, our findings reveal that one-particle excitations and
their corresponding dynamics can be faithfully represented
 only when a  Jastrow factor of a similar form is applied to
a coherent superposition of many Slater determinants. We discuss the
possible relevance  of this approach for the theoretical description of
photoemission spectra in higher dimensionality, where the present
wave function can be straightforwardly generalized
and  can be used as a variational ansatz, that is
exact for the 1D Luttinger model.
\end{abstract}

\maketitle

\section{Introduction}
Recently, much progress has been done for  understanding the crucial
role of strong electron correlation in  photoemission spectra,
namely the properties of the one-particle excitations and the
corresponding dynamical Green's function.\cite{shen2,shen3} This
topic  has attracted a   renewed attention due to the impressive
progress in the energy and momentum resolution  of  angle resolved
photoemission experiments, which confirm spin charge separation in
quasi one-dimensional systems\cite{shen1}, one of the most important
effects induced by strong electron correlation.

From the theoretical point of view, the pioneer work for the
dynamical Green's function dates to 1970\cite{br}, which showed the
absence of coherence in the strongly correlated regime of the
Hubbard model. Later the self-consistent
Born-approximation\cite{klr} was introduced, providing a
surprisingly  accurate description of the single hole dynamics
 in lattice models, such as the t-J model
 relevant for High-Tc superconductivity.
In this context, P.W. Anderson suggested that the Fermi liquid
picture could be violated not only in one dimension but also in
higher dimensionality (D), and especially in 2D. From this
speculation, a tremendous amount of work has been devoted to the
subject, starting from the reconsideration of the dynamical
properties of a single hole in a quantum
antiferromagnet\cite{poilblanc,siggia,sorellah}, and the detailed
analysis of the dynamical properties of a single minority spin
electron in a bath of fully polarized electrons with opposite
spin.\cite{edwards,castella,ssorellai,prok}

 Finally, an  important progress was achieved by the
Dynamical Mean Field Theory (DMFT), that was able to describe the
important feature of the Kondo resonance when approaching metal
insulator transition of the half-filled Hubbard model in infinite
dimensions.\cite{rmpkotliar,vollhardt} In this case it was also
shown that the single-particle excitations in the proximity of the
Mott transition may be highly nontrivial even in the metallic side.
In particular at  low energy, a Kondo resonance appears between the
expected upper and lower Hubbard bands and determines the coherent
quasi particle weight of the Fermi liquid metal, which vanishes
exactly at the metal-insulator transition. The predictions  of DMFT
have been confirmed by many experiments. For instance recently, Mott
transition in Vanadium Oxide was clearly
explained\cite{georgescience}. It represents also a theory capable
of characterizing the two energy scales found in photoemission
experiments of HTc compounds.\cite{dmftrec} However, the dynamical
properties predicted by this theory are not well understood outside
the DMFT formalism. In particular, it should be very important to
characterize the anomalous low-energy excitations  determining the
Kondo resonance from the direct solution of the Schrodinger
equation, namely  by direct inspection of the eigenfunctions of
strongly correlated models such as the Hubbard model or the t-J
model.

In this work, we consider a much less ambitious task and we use
  an  approach that can provide useful insights on the
exact ground state wave function and excitations. We consider the
well known
 Jastrow-Slater wave function that has been used successfully
in several correlated systems and we focus our analysis in one
dimension where analytic calculations are possible and numerical
works have confirmed the  impressive accuracy of the Jastrow-Slater
wave function on several strongly correlated
models\cite{mele,capello,capello1}. This wave function can be
generally  written as a product of a Slater determinant,
characterizing free electrons,
 times the so-called ``Jastrow factor" $J$
that appropriately weights the  electron configurations (e.g.
suppressing the wave function amplitude when the electrons are too
close), in order to describe electron correlation. Indeed, by a
lengthy but straightforward  derivation, we found that the exact
ground state of the Luttinger model can be written as a
Jastrow-Slater wave function.  Moreover, not only the ground state,
but also single-particle excitations can be written in a suitable
Jastrow-Slater form. It is interesting that in this case,
  {\em many}  Slater determinants have to be considered,
with an appropriate  change of the Jastrow factor. The results we
obtained  are in perfect agreement with the Luttinger liquid theory
in 1D and the extension of the wave function excitations to higher
dimensions seem to imply the same Kondo-like resonance scenario
obtained be means of DMFT, although at present this is just a
speculative conclusion and a further numerical work is necessary to
verify it.

This paper is organized as follows: In section \ref{Formalism}, we
review the Luttinger model and later show how the one-dimensional
fermionic Luttinger model Hamiltonian can be reduced to a quadratic
Bose Hamiltonian by means of the so-called bosonization technique.
In section \ref{BogoliubovSolving}, we show how the standard
Bogoliubov transformation can be used to diagonalize the quadratic
Bose Hamiltonian in order to obtain an exact ground state of the
Luttinger model. In section \ref{JastrowBCS}, we show that the
ground state of the Luttinger model can be rewritten as a Jastrow
wave function, whereas in section \ref{JastrowMSWF}, the
single-particle excitations of the Luttinger model  are given with
an explicit Jastrow multi-Slater wave function. Finally in section
\ref{dynamics}, we study the dynamics of a single particle added to
the right branch of the Fermi sea.

\section{Formalism}\label{Formalism}
\begin{figure}[ht]
\centering
\includegraphics[width=8cm, height= 6cm]{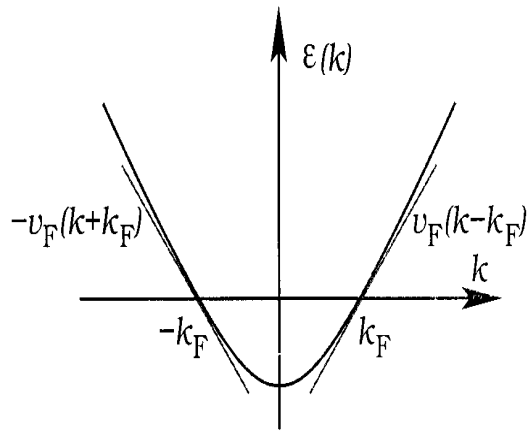}
\caption{Linearization of the realistic dispersion of a
one-dimensional Fermi system about the Fermi points $k_{F}$ and
$-k_{F}.$} \label{dispersion-1}
\end{figure}

 We consider the one-dimensional Luttinger model following the
work of Lieb and Mattis~\cite{LiebMattis} with the notations given
in a later work~\cite{SorellaParolaLutt}:
\begin{eqnarray} \label{luttinger}
H&=&v_F \sum_k  \left[( k -k_F) \psi^{\dag}_+(k) \psi_+(k)
 -( k +k_F) \psi^{\dag}_-(k) \psi_-(k)\right] \nonumber \\
 &+& {1 /L } \int\limits_0^L \int \limits_0^L dx ~dx~^\prime u(x-x^\prime) N(x)
N(x^\prime)~,
\end{eqnarray}
where $v_F$ is the Fermi velocity describing  a linear band around
the Fermi momentum $k_F$ and $u$ is a generic interaction which
depends only on the distance between electrons. In this
linearization scheme
 (see figure \ref{dispersion-1}), the allowed momenta in the right (+)
and left (-) branches
satisfy the usual  quantization conditions $k={ 2 \pi  \over L }  \times
{\it ~any~integer }$, valid for periodic boundary conditions, assumed here
and henceforth.
These two branches are then extended to $ -\infty , + \infty$
by means of  two fermionic fields
$\psi_{\pm }$ with their appropriate Fourier
transforms
\begin{equation}
\psi_{\pm} (k)
 ={ 1 \over \sqrt{L}} \int\limits_0^L dx~e^{ i k x } \psi_{\pm} (x)~.
\end{equation}

These fields define a local charge operator $N(x)= N_+(x)+ N_-(x)$
by means of the following contributions coming
 from the right and left branches:
\begin{equation}\label{N_di_x}
\left\{
\begin{array}{l}
N_{+}(x) = \psi_{+}^{\dagger}(x) \psi_{+}(x)
           - \langle \psi_{+}^{\dagger}(x) \psi_{+}(x) \rangle \\
 \\
N_{-}(x) = \psi_{-}^{\dagger}(x) \psi_{-}(x)
           - \langle \psi_{-}^{\dagger}(x) \psi_{-}(x) \rangle~.
\end{array} \right.
\end{equation}

The basic relations that make the Luttinger model exactly solvable
are given by the nontrivial commutation rules of these left and
right branches density operators\cite{LiebMattis}:
\begin{equation}\label{Commutatore_Nq}
\left[ N_{\pm}(q),~ N_{\pm}(-q) \right] = \mp \frac{Lq}{2 \pi}~,
\end{equation}
where
\begin{equation}\label{Nq_Fourier}
N_{\pm}(q) = \int \limits_0^L dx e^{-iqx} N_{\pm}(x)~.
\end{equation}

After this important observation, it is possible to represent these
density operators in terms of canonically conjugate  bosonic fields
$\Phi$ and $\Pi$\cite{affleck}:
\begin{equation}\label{Bosonizz1}
\left\{
\begin{array}{l}
N_{+}(x) =  \frac{1}{\sqrt{4\pi}} \left[ \Pi(x) + \partial_{x}\Phi(x) \right] \\
 \\
N_{-}(x) =  -\frac{1}{\sqrt{4 \pi}} \left[ \Pi(x) -
\partial_{x}\Phi(x) \right]
\end{array} \right.
\end{equation}
with
\begin{equation}\label{Commutatore_PHI_PI}
\left[ \Phi(x), ~\Pi(x')  \right] = i \delta(x-x')~.
\end{equation}
The total density operator $N(x)$ can be written as

\begin{equation}\label{Nx_Phi_x}
N(x)= N_{+}(x) + N_{-}(x) =  \frac{1}{\sqrt{\pi}}
\partial_{x}\Phi(x)~.
\end{equation}

Now it takes just a little more algebra to show that the fermionic
Luttinger Hamiltonian (\ref{luttinger}) can be expressed in terms of
these bosonic fields where it assumes the following canonical
harmonic form:

\begin{equation}\label{hamLutBosx}
H = \frac{v_{s}}{2} \int_{0}^{L} dx \left\{ K {\Pi}^2(x) +
\frac{1}{K} \left[  \partial_{x} \Phi(x) \right]^{2} \right\}~,
\end{equation}
where $v_{s}$ is the re-normalized Fermi velocity and $K$ is the
Luttinger parameter that can be simply expressed in terms of  $v_F$
and the $k=0$ component of the interaction
$u$.\cite{LiebMattis,LederLut}. In principle, the mapping of Eq.
(\ref{luttinger}) to the harmonic Hamiltonian (\ref{hamLutBosx}) is
exact only  if the interaction is assumed to be delta-like, but it
can be easily generalized to a momentum-dependent coupling constant
$u$, as it was done for instance in Ref.\cite{LiebMattis}.
 This canonical form given in Eq. (\ref{hamLutBosx}) embodies the
entire low-energy physics of the so-called  Luttinger liquids and
due to its simplicity, can be solved explicitly. To this purpose, we
introduce Fourier transforms of the bosonic fields:
\begin{equation}\label{PHI_PI_Fourier}
\left\{
\begin{array}{lll}
\Phi_{k} ={1 \over \sqrt{L} }
 \int_{0}^{L} dx \Phi(x) e^{-ikx} & \leftrightarrow &
\Phi(x)  =  {1 \over \sqrt{L} } \sum_{k} \Phi_{k} e^{ikx}   \\
 \\
\Pi_{k} = {1 \over \sqrt{L} }
 \int_{0}^{L} dx \Pi(x) e^{ikx} & \leftrightarrow &
\Pi(x)  =  {1 \over \sqrt{L} } \sum_{k} \Pi_{k} e^{-ikx}   \\
\end{array} \right.
\end{equation}

By making use of the relations in (\ref{PHI_PI_Fourier}),
~(\ref{hamLutBosx}) becomes:

\begin{equation}\label{hamLutBosq}
H = \frac{v_{s}}{2} \sum_{k} \left[
 K \left( \Pi_{k} \Pi_{-k} \right)
+ \frac{1}{K}  \left( k^2 \Phi_{k} \Phi_{-k} \right) \right]~,
\end{equation}
which expresses the Hamiltonian in terms of ``normal coordinates",
$\Phi_{k}$ and $\Pi_{k}$. Notice that we have essentially reduced
the problem to a single harmonic oscillator for each given momentum.
The next step merely repeats the procedure carried out for a single
harmonic oscillator. We define a set of conjugate creation and
annihilation operators:

\begin{equation}\label{Bosonic_Operator}
\left\{
\begin{array}{lll}
\Phi_{k}   & = &
 \frac{1}{\sqrt{2 |k| }} \left[ a_{k}^{\dagger} + a_{-k} \right]   \\
 \\
\Pi_{k}  & = &
i \sqrt{ \frac{|k|}{2} } \left[ a_{-k}^{\dagger} - a_{k} \right]~.   \\
\end{array} \right.
\end{equation}

Now substituting (\ref{Bosonic_Operator}) into (\ref{hamLutBosq}),
we obtain the standard quadratic Bose Hamiltonian:
\begin{equation}\label{hamLutBCS}
H = \sum_{k} \epsilon_{k} \left[  a_{k}^{\dagger} a_{k} +
\frac{1}{2} \right] - \sum_{k} \frac{\gamma_{k}}{2} \left[
a_{k}^{\dagger} a_{-k}^{\dagger} + a_{k} a_{-k} \right]~,
\end{equation}
where the functions $\gamma_k$ and $\epsilon_k$  are given by:

\begin{equation}\label{ParametriBCS}
\left\{
\begin{array}{lll}
\epsilon_{k}  & = &
\frac{v_{s}|k|}{2} \left[ K + \frac{1}{K}  \right] \\
 \\
\gamma_{k}  & = &
\frac{v_{s}|k|}{2} \left[ K - \frac{1}{K}  \right] ~.\\
\end{array} \right.
\end{equation}
The Hamiltonian (\ref{hamLutBCS}) contains only excitations in a
given sector of particle number because in this section, we are
interested in determining only the ground state (GS).

\section{Exact solution of the model by Bogoliubov transformation}\label{BogoliubovSolving}

We will use the following unitary transformation which is the
standard Bogoliubov transformation to diagonalize Eq.
(\ref{hamLutBCS}):
\begin{equation} \label{eis}
U  = e^{ i S }~,
\end{equation}
where $S$ has been obtained in Ref. (\onlinecite{LiebMattis}):
$$S=i \sum_{q>0} \theta_q  a^{\dag}_q a^{\dag}_{-q} + {\rm h.c. } ~.$$
It is easy to show that
\begin{equation} \label{BosTrans_S}
U a_q U^{\dag}  = cosh(\theta_q) a_q +sinh(\theta_q )
a^{\dag}_{-q}~.
\end{equation}
It follows immediately that the transformed Hamiltonian $ H \to U H
U^{\dag}$ assumes the diagonal form

\begin{equation}\label{hamdiagbog}
 UHU^{\dag} = \sum_{k| E_k >0 }\left( E_{k} a_{k}^{\dagger} a_{k} +
\frac{\epsilon_{k}}{2}\right)~,
\end{equation}
where
\begin{equation}\label{SolCalc2}
E_{k} = \sqrt{ \epsilon_{k}^2 - \gamma_{k}^2 } = v_s |k|~.
\end{equation}
This can only be possible provided
\begin{equation}\label{relazioneUkVk}
\tanh (\theta_k ) =- \frac{\gamma_{k}}{ \epsilon_{k} +
\sqrt{\epsilon_{k}^2 - \gamma_{k}^2} }~,
\end{equation}
 namely that $\theta_k$ does not depend on the momentum
$k$ and is simply given in terms of the Luttinger parameter $K$,
$$e^{ -2 \theta_k} = K~.$$

After this unitary transformation,
 the ground state cannot contain any boson excitation
because $E_k>0$, implying that in this representation, the ground
state  coincides with the vacuum $|0\rangle$, namely the
non-interacting Fermi sea $|FS\rangle$, the unique state of the
Luttinger model corresponding  to the vacuum of the canonical
operators $a_{k}$ ($a_{k}|FS\rangle = 0,~\forall k$).

It follows that in  the original representation, the GS of the Bose
Hamiltonian is simply  given by $U |FS \rangle$ and can be written
in the following general form up to an irrelevant normalization
constant:
\begin{equation}\label{groundstateBCSansatz}
| GS \rangle = exp \left\{
 \sum_{q>0} f_{q} a_{q}^{\dagger} a_{-q}^{\dagger} \right\} | FS
 \rangle~.
\end{equation}
Here, $f_q$ is a function whose analytic form will be determined in
what follows. In order to determine $f_q$, we notice that $U a_{k}
U^{\dag}$ must annihilate the GS $U | FS \rangle$
  $\forall
k$. After substituting $U | FS \rangle$ with the ansatz
(\ref{groundstateBCSansatz}) and by means of Eq. (\ref{BosTrans_S}),
we obtain a simple equation
\begin{equation}\label{solContoPairing}
 a^{\dag}_{-k} ( \cosh(\theta_{k})  f_{k} + sinh(\theta_{k})) = 0
\end{equation}
which can be solved to get
\begin{equation}\label{finalpairingBCS}
f_{k} = -\tanh(\theta_{k})  = \frac{\gamma_{k}}{ \epsilon_{k} +
\sqrt{\epsilon_{k}^2 - \gamma_{k}^2} }~.
\end{equation}

By replacing  the quantities $\epsilon_{k}$ and $\gamma_{k}$
from~(\ref{ParametriBCS}) into~(\ref{finalpairingBCS}), we obtain
the pairing function in terms of the Luttinger interaction parameter
$K$:

\begin{equation}\label{LuttinPairingBCS}
f_{k} = { \sqrt{K}  -{1 \over \sqrt{K}}   \over
 \sqrt{K} +{1\over \sqrt{K }}  }~.
\end{equation}
When $K = 1$ (i.e. in the non-interacting regime), the pairing
function $f_{k}$ correctly vanishes, and the ground state is just
the Fermi sea $|FS\rangle$.


\section{ground state of the Luttinger model by the
Jastrow-Slater wave function}\label{JastrowBCS}

In this section, we show that the ground state of the Luttinger
model is a Jastrow-Slater wave function i.e. the ground state
(\ref{groundstateBCSansatz}) can be rewritten as a Jastrow wave
function $|\psi_J \rangle$:
\begin{equation}
|\psi_J\rangle = e^{- {1 \over 2} \sum_q v_q N_q N_{-q} }  |FS
\rangle~,
\end{equation}
where $|FS\rangle$ is the free Fermi sea,
\begin{equation}\label{Nk_Phi_k_Jas}
N_{q} = i { q \over \sqrt{\pi} } \Phi_{q}=i q \sqrt{ { 1 \over 2 \pi
|q|}}  ( a^{\dag}_q + a_{-q} )~,
\end{equation}
and $v_q $ are the momentum-dependent parameters.

Henceforth we assume that $v_q\ge  0$ and that $v_q=0$ for $q=0$,
because the total charge $ \int\limits_0^L dx ~N(x)$ is conserved.
 In this way  we will  obtain the analytic form of the Jastrow
parameters ($v_{q}$) as functions of the pairing amplitude $f_q$, by
requiring that
\begin{equation}\label{startingpoint}
e^{ -\left\{ \sum_{q > 0} v_{q} N_{q} N_{-q} \right\} } |FS \rangle
= R_\alpha e^{ \left\{ \sum_{k>0} f_{k} a_{k}^{\dagger}
a_{-k}^{\dagger} \right\} } |FS \rangle~,
\end{equation}
where $R_\alpha$  is an overall constant which depends only on
$f_{q}$. In the following derivation we do not even need to assume
that the pairing function $f_k$ is constant.

Let us first introduce the Hubbard Stratonovich
transformation for the Jastrow factor:
\begin{equation}\label{hst}
e^{- \left\{ \sum_{q > 0} v_{q} N_{q} N_{-q} \right\} } = \int
\prod_{q>0}   { dz_q \over \pi}  e^{ - |z_q|^2 +
 \sqrt{v_q} ( z_q N_q - z^*_q N_{-q} )  }~.
\end{equation}
Using Eq. (\ref{Nk_Phi_k_Jas}), the RHS of the above equation
simplifies to
\begin{equation} \label{hst_1}
e^{- \left\{ \sum_{q > 0} v_{q} N_{q} N_{-q} \right\} } = \int
\prod_{q>0}  { dz_q \over \pi} e^{ - |z_q|^2  +  A_q+ B_q }~,
\end{equation}
where
\begin{eqnarray}\label{AandB_Jast}
A_q  & = & i q \sqrt{{ v_q \over 2  \pi |q|} } ( a^\dag_{q} z_q
+ a^\dag_{-q}  z^*_q) \\
B_q  & = & i q \sqrt{{ v_q \over 2  \pi |q|} } ( a_{-q} z_q + a_q
z^*_q)~.
\end{eqnarray}

Now  using the Baker Haussdorf Campbell formula $ e^{ A_q +B_q } =
e^{ A_q } e^{ B_q} e^{ -1/2 [ A_q,B_q ] } $
 (valid if $[A_q,[A_q,B_q]]= [B_q,[A_q,B_q]] =0$
as in this case), the commutator in the previous expression can be
explicitly evaluated and it is a constant:
\begin{equation}\label{calc_commutatBHC}
- \frac{1}{2} \left[ A_q , B_q \right] = -\frac{ |v_{q}| |q|}{ 2 \pi
} |z_q|^2~.
\end{equation}
On the other hand, $e^{ B_q} |FS\rangle= |FS\rangle $ because all
non-vanishing powers of $B_q$ annihilate the vacuum. Thus, we obtain
the following equation after applying  the  operator
 (\ref{hst_1}) to $|FS\rangle$:

\begin{equation}\label{calc2_commutatBHC}
e^{- \left\{ \sum_{q > 0} v_{q} N_{q} N_{-q} \right\} }|FS \rangle =
\int \prod_{q>0}  { dz_q \over \pi}  e^{- |z_q|^2 ( 1+ \alpha_q   )
}  e^{ A_q } | FS \rangle~,
\end{equation}
where
\begin{equation}
\alpha_q= { |q| v_q  \over 2 \pi } >0~.
\end{equation}

By performing the remaining
 simple Hubbard-Statonovich transformation integral we obtain:

\begin{equation}\label{calc3_commutatBHC}
|\psi_J \rangle  = R_{\alpha}
    e^{ -\sum_{q>0}  {  \alpha_q \over 1+ \alpha_q } a^{\dag}_q a^{\dag}_{-q} }
   | FS \rangle~,
\nonumber
\end{equation}
where
\begin{equation} \label{Rconst}
R_{\alpha} = \prod_{q>0} \frac{1}{  1+\alpha_{q} }~.
\end{equation}
Notice that this constant can be infinite if $v_q$ does not decay
sufficiently fast for large $q$. The divergence in the infinite
product can generally be removed by introducing a  large momentum
cutoff, e.g. $|q| < \Lambda_{cut}$, and taking into account that
$R_\alpha$ is just an overall normalization constant that does not
change any physical expectation value even when the  cutoff is sent
to infinity.

Now by a direct comparison with Eq. (\ref{startingpoint}), we obtain
that the Jastrow wave function $\psi_J$ is the ground state of the
Luttinger model if and only if
\begin{equation}
{\alpha_q \over 1+ \alpha_q } = -f_q= { {1 \over \sqrt{K}} -\sqrt{K}
\over
 \sqrt{K} +{1\over \sqrt{K} } }~.
\end{equation}
It follows immediately from the above equation that for $K < 1$, the
ground state momentum-dependence of the Jastrow
parameters can be expressed as
\begin{equation} \label{vqdef}
v_q  = {\pi \over |q|} \left({ 1\over K} -1\right)~.
\end{equation}

\subsection*{Remark}

From Eq. (\ref{vqdef}), it follows immediately that when $K = 1$
(corresponding to the free theory), $v_q  = 0$  $\forall$ q and
hence $|\psi_J \rangle $ reduces to $|FS\rangle$ which is the ground
state of the free theory.


\section{Single-particle excitations by a Jastrow Multi-Slater wave function}\label{JastrowMSWF}
In general, not only the ground state but all the  eigenstates of
the Luttinger Hamiltonian can be written in a Jastrow multi-Slater
form with appropriate Jastrow factors. To this purpose, let us
consider that an eigenstate of the non-interacting Luttinger model $
|FS_k \rangle$ (defined later) can be transformed onto an exact
eigenstate of the interacting  Luttinger model ($K\ne 1$) by means
of the unitary transformation $U=e^{i S}$ (defined in the previous
chapter in Eq. (\ref{eis})),  namely
\begin{equation} \label{exk}
|\psi_k \rangle  = e^{ i S } |FS_k \rangle~.
\end{equation}
Here, $|FS_k\rangle$ is a suitable excited state in the free theory
with an extra particle added to the right branch slightly above the
Fermi momentum (as we are interested in low-energy excitations
only). This excited state can be expressed as
\begin{equation}
| FS_k\rangle =  \int\limits_0^L ~dx~ e^{-i k x } \psi^{\dag}_{+}
(x) |FS\rangle~.
\end{equation}
The right-moving operator $\psi^{\dag}_+(x)$ after integration over
$x$, creates a state, namely a non-interacting Slater determinant
 with an extra particle added to the right branch and with total  momentum $k+k_F$.

The eigenstate $|\psi_k\rangle $ defined in (\ref{exk}) can be
expressed exactly in the following generalized Jastrow-Slater form
(see appendix \ref{singpartex} for a detailed derivation):
\begin{eqnarray} \label{psijh}
|\psi_{J,h} \rangle &=&
 \int\limits_0^L dx ~~  {\rm exp}
\left[ - {1 \over 2} \sum_q v_q N_q N_{-q} \right]\times \\
  &  \times&{\rm exp}  \left[  \sum_q h_q e^{ i q x } N_q  \right]
 \psi^{\dag}_+(x) e^{-i k x } |FS \rangle~. \nonumber
\end{eqnarray}

\subsection{Remark}
In appendix \ref{singpartex}, we establish the dependence of $h_{q}$
on the interaction parameter $K$. It can be shown that when $K = 1$
(corresponding to the free theory), $v_{q} = 0$, $h_{q} = 0$ and the
excited state $\psi_{J,h}$ reduces to $| FS_k\rangle$ which is a
single determinant. Generally, when $K\neq1$, $h_{q}\neq 0 $ and in
this case the excited state $\psi_{J,h}$ represents a
multi-determinant Jastrow correlated state, very similar to the
Edward's ansatz used for the single spin flip state of the
ferromagnetic Hubbard model described in the
introduction\cite{edwards}. Our derivation shows that
 this ansatz is exact for the
Luttinger model.
 Strictly speaking, the number of
determinants required to describe the ansatz is infinite because in
Eq. (\ref{psijh}), the variable $x$ is continuous and a Slater
determinant is required in general for each value of $x$. In
practice however, we can go back to a lattice discretized version of
the Luttinger  model
 where the position of an electron is discretized and therefore
for generic lattice models, the  number of determinants required for
describing the present ansatz should scale as the number of lattice
sites. In this respect, our result does not simply mean that
one-particle excitations of the Luttinger liquid can be expressed as
linear superpositions of Slater determinants (this implication would
be
 trivial because any state can be expressed in this way),
but provides an important restriction to the form of the wave
function because the
 number of Slater determinants used in this ansatz remains much smaller
than the dimension of the Hilbert space.

\subsection{Spectral property of the Luttinger liquid: the quasiparticle
weight}\label{spectral} With this formalism, one can recover most
of
 the exact results obtained with  the conventional
bosonization technique  since we have represented the ground state
and the one-particle excitations using a different (but equivalent)
functional form. For instance, we evaluated the quasiparticle weight
for the Luttinger liquid determined by the ground state and the
lowest one-particle excitation. Our calculation obtained with
Jastrow-Slater wave functions implies that the quasiparticle weight
vanishes  in the thermodynamic limit according to the power law $Z
\sim L^{-\theta}$, where $\theta=(K + K^{-1}-2)/2$. This result ties
in perfectly with Luttinger liquid theory\cite{LiebMattis, voit}.

\section{One-particle  Dynamics}\label{dynamics}

The above result can be extended to represent the quantum mechanical
time evolution $e^{ i H t}$ of any one-particle state in a suitable
time-dependent
 Jastrow-Slater form.

In particular, we apply to the ground state written as  $e^{ i S}
|FS \rangle$, the operator $\int\limits_0^{L}dx~e^{-i k x }
\psi^{\dag}_{+} (x)$ that creates a fermion with momentum $k+k_F$ in
the right branch. In this section we determine the exact time
evolved state
\begin{equation}\label{holeGSstart}
| \Psi (t,k) \rangle = e^{ i H t }\int\limits_0^L  ~dx ~e^{-i k x }
\psi^{\dag}_{+}  (x) e^{ i S }  |FS \rangle
\end{equation}
in terms of a generalized Jastrow-Slater form. This form is similar
to what we obtained for the one-particle excitation $|\psi_k \rangle
$ in the previous section. Since all  the single-particle
excitations have the same two-body Jastrow potential $v_q$, it is
reasonable to expect the following ansatz
\begin{eqnarray}
| \Psi_{JhC} (k,t) \rangle &=&
 \int\limits_0^L dx ~C(t)~ e^{-i k x }
 {\rm exp}
\left[ - {1 \over 2} \sum_q v_q N_q N_{-q} \right] \times\nonumber \\
& \times&{\rm exp}  \left[  \sum_q h_q(t)  e^{ i q x}  N_q  \right]
 \psi^{\dag}_+ (x)  |FS \rangle ~.\nonumber \\
\end{eqnarray}

Indeed this is the exact time-evolved state within the Luttinger
model Hamiltonian provided the functions $h_q(t)$ are chosen
appropriately ( see appendix \ref{dynder}).

\section{Discussion}

The most important outcome of this work  is that it is possible to
obtain an essentially exact description of the low-energy properties
of one-dimensional correlated models by means of a Jastrow-Slater
wave function. Indeed, we have shown that the ground state of the
Luttinger model; the well known and accepted model for describing
low-energy physics in one dimension, can be written exactly as a
long-range
 Jastrow factor applied to the uncorrelated Fermi sea.

In addition, not only the ground state of the Luttinger model but
also single-particle excitations can be described by the Jastrow
wave function. In this case however, the Jastrow factor is applied
to many Slater determinants. Thus, unlike the non-degenerate ground
state which is just a Jastrow-Slater single determinant, the excited
state has an intrinsic multi-determinant character. This
multi-determinant state reflects the effect  of inserting a single
particle to the ground state of the Luttinger model. The wave
function for the system with an extra particle changes drastically
in form  within the Jastrow-Slater ansatz i.e., from a single
determinant to a multi-determinant state. The dynamical properties
of the Luttinger liquid were also formulated within the
Jastrow-Slater wave function. More specifically, the dynamics of a
single particle added to the ground state of the Luttinger model can
be expressed  as a Jastrow multi-Slater state, with time-dependent
and complex Jastrow factors.

To avoid confusion, we remark  here that this is just a new
alternative point of view of the well known exact solution of the
Luttinger model in one dimension. However, it is important to
emphasize that the Jastrow-Slater wave function approach can be
easily extended to higher dimensionality and indeed, a long-range
Jastrow factor applied to a Slater determinant has been used widely
for electronic simulations based on the so-called quantum Monte
Carlo technique \cite{Alder}. However, the present form of a
multi-determinant Jastrow factor, that we have shown to be a
necessary ingredient to deal with the single-particle excitation
spectrum of the Luttinger model, has never been used before to the
best of our knowledge. This form is particularly important in one
dimension, in order to destroy the quasiparticle weight and
determine the non-Fermi liquid behavior. Therefore, we expect that
the extension of this wave function to higher dimensionality  may
lead to a deeper understanding of the photoemission spectrum of
strongly correlated materials like high temperature superconductors
which display unconventional behavior \cite{Yunoki,Norman}. In these
systems, the photoemission spectrum is still controversial and
unexplained. For instance, the strong momentum dependence in angle
resolved photoemission experiments \cite{Mnorman,Nicolas,shen} with
Fermi arcs or hole pockets cannot be easily reproduced with the
conventional single determinant Jastrow-Slater ansatz for the single
particle excitations.\cite{Yunoki} On the contrary, the extension of
the wave function defined in Eq. (\ref{psijh}) to higher
dimensionality provides  a variational ansatz containing more
variational freedom for the excitations (e.g. $h_q$ in Eq.
\ref{psijh}), that may lead to more accurate results and possibly
better agreement with experiments. Obviously, a systematic
variational Monte Carlo study outside the scope of this work is
necessary to confirm this interesting possibility.

 Finally, we would
like to comment on the possible explanation of the Kondo resonance
in the spectral weight of a metal, predicted by DMFT in infinite
dimensions. In our approach, the wave function of an added particle
with momentum $k$ can be viewed as  a coherent superposition of
Slater determinants with a real space defect located at each space
position $x$ (see Eq. (\ref{psijh})). The excitation  described in
Eq. (\ref{psijh}) is very similar to old types of wave
functions\cite{edwards} and previous approaches to consider the
single hole dynamics\cite{SorellaParolaLutt}. The common feature of
this approach with  the Kondo problem is that the single-particle
(or -hole) excitation acts like a real space impurity in the frame
where the extra particle is taken fixed. The impurity problem is a
peculiar characteristic of the Kondo model fixed point and
therefore, we expect that the Jastrow-Slater approach should be able
to introduce another energy scale, the Kondo one, in the problem of
the photoemission spectrum, very similarly to the  DMFT scenario.
Again also in this case,  numerical work is necessary to  verify
this issue.

\acknowledgments We acknowledge useful discussions with M. Lugas,
at the early stage of this work.

\newpage
\appendix

\section{ Detailed derivation of the single-particle excited state}  \label{singpartex}
We report below the detailed derivation of the single-particle
excited wave function. Although it may appear cumbersome and
elaborated, it is indeed very simple conceptually. All one-particle
excitations, as well as the ansatz state (\ref{psijh}), can always
be normal ordered according to the obvious rule that:
\begin{itemize}
\item the fermionic operator $\psi^{\dag}_+ (x)$ is always the leftmost one.
\item after that all bosonic terms
can be ordered in the normal way: the creation operators
$a^{\dag}_q$  to the left and the destructions $a_q$ to the right
positions.
\item In this way all the destruction operators $a_q$ disappear
because they have to be applied to the vacuum and the final
expression drastically simplifies.
\end{itemize}

Indeed after the above three steps, it is easy to convince ourselves
that one can generally obtain,  both from the exact  excitation
Eq.(\ref{exk}) and the ansatz Eq.(\ref{psijh}), similar expressions
containing, after  the fermionic operator,
 a gaussian form of operators $a^{\dag}_q$ applied to the
vacuum. In this way,  it is possible to match the two expressions.

\subsection{First step, simplification of the ansatz $|\psi_k \rangle$ }

In order to apply  $e^{i S} $
 to the fermion  field $\psi^{\dag}_{+} (x)$
 we follow Ref.(\onlinecite{LiebMattis})
that have derived that:
\begin{equation} \label{LM}
e^{ i S } \psi^{\dag}_{+} (x) e^{-i S } = \psi^{\dag}_{+} (x) {\rm
exp} \left[\sum_q e^{i q x } ( h^{LM}_q   a^{\dag}_q +  h^{LM}_{-q}
a_{-q} ) \right]
\end{equation}
where for  $ q > 0$,
\begin{equation} \label{h_LM_expression}
\left\{
\begin{array}{l}
h^{LM}_q =  i \sqrt{ { 2 \pi \over |q | L }
}(\cosh~\theta_{q}-1) \\
 \\
h^{LM}_{-q}=  i \sqrt{ { 2 \pi \over |q | L } }\sinh~\theta_{q}~.
\end{array} \right.
\end{equation}
We can use  Eq.(\ref{LM}) to simplify the exact excitation $|\psi_k
\rangle $ given by  Eq.(\ref{exk}):
\begin{eqnarray}
&|\psi_k\rangle &= \int\limits_0^L dx ~ e^{ - i k x}
 e^{ i S} \psi^{\dag}_{+}(x) e^{-i S}  e^{iS} |FS \rangle = \\
  & \int\limits_0^L dx & \psi^{\dag}_{+}(x)
{ \rm exp  } \left[ \sum_q  e^{i q x } ( h_q^{LM} a^{\dag}_q   +
h_{-q}^{LM}  a_{-q}  )
 \right]
e^{ i S } |FS\rangle \nonumber
\end{eqnarray}

We recall that the ground state of the Luttinger model, apart for a
normalization constant, is given by:
\begin{equation}
|GS\rangle = e^{ F } | FS \rangle
\end{equation}
where:
\begin{equation}\label{defF}
F=\sum_{q>0 } f_q  a^{\dag}_{q} a^{\dag}_{-q}.
\end{equation}
In this way  we obtain:
\begin{eqnarray}
\lefteqn{|\psi_k\rangle= R_\alpha \times} \nonumber \\
&&\int\limits_0^L dx \psi^{\dag}_{+}(x) e^{F } { \rm exp  } \left[
\sum_q e^{i q x }  e^{-F} (h^{LM}_q a^{\dag}_q + h^{LM}_{-q} a_{-q}
) e^{F}   \right] |FS\rangle\nonumber
\end{eqnarray}
Finally by using
\begin{eqnarray} \label{applyF}
e^{-F}  a_{p} e^{F }&=& f_{p}  a^{\dag}_{-p} +a_{p} \nonumber \\
e^{-F}  a^{\dag}_{p} e^{F }&=&  a^{\dag}_{p}
\end{eqnarray}
and applying again $ e^{ A +B } = e^{ A } e^{ B } e^{ -1/2 [ A,B ] }
$ with $A=\sum_q (h^{LM}_q +f_q h^{LM}_{-q}) e^{i q x}
 a^{\dag}_q $
and $B=\sum_q h_q^{LM} e^{-iq x }  a_{q}$ we obtain:

\begin{eqnarray} \label{basic}
|\psi_k \rangle &=&  C \int\limits_0^L~ e^{ - i k x} dx
~\psi^{\dag}_{+}(x) e^{F} \times \nonumber\\
&\times&{\rm exp} \left[\sum_q (h_q^{LM}+f_q
h_{-q}^{LM} ) a^{\dag}_q e^{ i q x}  \right] |FS \rangle ~,\\
 C &=&
{\rm exp} \left( 1/2 \sum_q h_q^{LM} (h_{q}^{LM}+ f_q h_{-q}^{LM})
\right)~. \nonumber
\end{eqnarray}

\subsection{ Second step:
 simplification of the ansatz $ |\psi_{Jh} \rangle $ }
\label{subsec:Jh} Also the ansatz $\psi_{LJ} $ can be recast in a
form similar to Eq.(\ref{basic}) using even simpler algebra because
the Jastrow depends only on the total density operator $N_q$,
with   commutation rules
\begin{eqnarray}
\left[ N_q, N_{q^\prime}\right] &=& 0 \\
\left[ N_q, \psi^{\dag}_+(x) \right] &=&  { 1 \over \sqrt{L} } e^{ -
i q x } \psi^{\dag}_+(x).
\end{eqnarray}
In this way it is easy  to derive the following
useful relations:

\begin{eqnarray*}
e^{ -1/2 \sum_{q} v_q N_q N_{-q} }& \psi^{\dag}_+(x)&
e^{ 1/2 \sum_{q} v_q N_q N_{-q} }\\
 = &\psi^{\dag}_+(x) &
{\rm exp} -\left[ \sum_{q} { v_q \over \sqrt{L} }   ( e^{i q x } N_q
+
{1 \over 2\sqrt{L} } ) \right]  \\
e ^{\sum_q h_q N_q e^{i q x } } &\psi^{\dag}_+(x)&
e ^{-\sum_q h_q N_q e^{i q x } } \\
 =
&\psi^{\dag}_+(x) & {\rm exp } \left[ \sum_q { h_q \over \sqrt{L}}
\right]
\end{eqnarray*}
With the above relations we can bring the operator
 $\psi_(x) = e^{i k_F x}  \psi_+^{\dag} (x) $ (as
all the left branch states are occupied for momenta $p\simeq k_F$ as
we assume) in the leftmost side of Eq.(\ref{psijh}) and obtain:

\begin{eqnarray}  \label{psijh1}
\nonumber |\psi_{J,h} \rangle &=& {\rm exp} \left[ \sum_q \left({
z_q^{(1)} \over \sqrt{L}} \right)\right] \times  \\\nonumber
&\times& \int\limits_0^L dx ~  \psi^{\dag}_+(x) e^{-i k x }
    {\rm exp}  \left[  \sum_q   z_q^{(2)}
e^{ i q x } N_q  \right] \times \\
 &\times& {\rm exp}  \left[ - {1 \over 2}
\sum_q v_q N_q N_{-q} \right]
  |FS \rangle
\end{eqnarray}
where
\begin{equation} \label{defzq1}
z_q^{(1)}= h_q- { v_q \over 2\sqrt{L}}
\end{equation}
and
\begin{equation} \label{defzq}
z_q^{(2)}= h_q- { v_q \over \sqrt{L}}.
\end{equation}

Our ansatz can be further simplified by implementing the condition
which enabled us to obtain the Jastrow parameter in terms of the
pairing function, namely :

\begin{equation}\label{step2-7}
 {\rm exp}  \left[ - {1 \over 2} \sum_q v_q N_q N_{-q} \right]
  |FS \rangle = R_{\alpha}  e^{F} |FS \rangle,
\end{equation}
where $F$ has been previously defined in Eq.(\ref{defF}). Now we can
replace $N(q)$ in terms of canonical operators as in
Eq.(\ref{Nk_Phi_k_Jas}) and we can perform similar steps as before,
namely:

\begin{eqnarray}\label{step2-8}
|\psi_{Jh} \rangle &=&
 R_{\alpha}
{\rm exp} \left[ \sum_q ({ z_q^{(1)}  \over \sqrt{L}} ) \right]
 \int\limits_0^L dx ~  \psi^{\dag}_+(x) e^{-i k x }\times \nonumber \\
& \times& {\rm exp} \left[ \sum_q { i q z_q^{(2)} e^{ i q x} \over
\sqrt{ 2 \pi |q| }  } (a^{\dag}_q + a_{-q} ) \right] e^F|FS\rangle
\nonumber  \\\nonumber.
\end{eqnarray}


This can be also written as

\begin{eqnarray}\label{step2-9}
|\psi_{Jh} \rangle &=&
 R_{\alpha}
{\rm exp} \left[ \sum_q ({ z_q^{(1)}  \over \sqrt{L}} ) \right]
 \int\limits_0^L dx ~  \psi^{\dag}_+(x) e^{-i k x }\nonumber \times \\
&\times& e^F {\rm exp} \left[ \sum_q { i q z_q^{(2)} e^{ i q x}
\over \sqrt{ 2 \pi |q| }  }  e^{-F} (a^{\dag}_q + a_{-q} ) e^F
\right] |FS\rangle \nonumber  \\\nonumber.
\end{eqnarray}

Finally, using the relations in eq.(\ref{applyF}) and the fact that
$a_{q}|FS\rangle = 0$ $\forall q$, we obtain:

\begin{eqnarray} \label{finalJh}
|\psi_{Jh} \rangle &=& C_{Jh}
 \int\limits_0^L dx ~  \psi^{\dag}_+(x) e^{-i k x }e^F \times \nonumber\\
& \times&{\rm exp}
 \left[ \sum_q { i q  z_q^{(2)} (1+ f_q) e^{ i q x}
 \over \sqrt{ 2 \pi |q|} } a^{\dag}_q \right]
|FS \rangle
\end{eqnarray}
where

\begin{eqnarray*}
C_{Jh} &=& R_{\alpha} {\rm exp } \left[ \sum_q ( { z_q^{(1)} \over
\sqrt{L} } -{ |q|   \over 4 \pi } z_q^{(2)} z_{-q}^{(2)} (1 + f_q)
\right]\nonumber.
\end{eqnarray*}

 We are now in the position to match the two states $|\psi_{Jh} \rangle $ and
$|\psi_k\rangle $,  using equations (\ref{finalJh}) and
(\ref{basic}) respectively. Indeed,  apart from an irrelevant constant,
the ansatz $|\psi_{Jh} \rangle $ is an exact excited state of the
Luttinger model, if the following condition is satisfied:

\begin{equation} \label{finalrel}
{ i q z_q^{(2)} (1+f_q) \over \sqrt{2 \pi |q|}  } = h_q^{LM} +f_q
h_{-q}^{LM}.
\end{equation}
Since $z_q^{(2)}$ is linear in $h_{q}$, the above equation is a
simple linear equation, that can be solved for the unknown quantity
$h_{q}$.
\section{ Detailed derivation of the real time evolution for the Jastrow Slater
wave function}  \label{dynder}
\subsection{Simplification  of the state $ \Psi_{JhC} $ }
This    can be obtained by applying the same derivation described in
the subsection (\ref{subsec:Jh}), with slightly different notations.
Therefore the final expression is:
\begin{eqnarray} \label{finalJhC}
|\Psi_{JhC} \rangle &=&
 \int\limits_0^L dx  e^{-i k x }   \psi^{\dag}_{+} (x  ) C_{Jh} (t)
  C(t) \times \nonumber\\
& \times& e^F{\rm exp}
 \left[ \sum_q { i q  z_q^{(2)} (1+ f_q) e^{ i q x}
 \over \sqrt{ 2 \pi |q|} } a^{\dag}_q \right]
|FS \rangle \\
C_{Jh} (t)  &=& R_{\alpha} {\rm exp } \left[ \sum_q ( { z_q^{(1)}
\over \sqrt{L} } -{ |q|   \over 4 \pi } z_q^{(2)} z_{-q}^{(2)} (1 +
f_q) \right]~.\nonumber
\end{eqnarray}
Here,
\begin{equation}
z_q^{(2)} (t) = h_q(t) -{ v_q \over \sqrt{L}}
\end{equation}
and
\begin{equation}
z_q^{(1)} (t) = h_q(t) -{ v_q \over 2\sqrt{L}}~.
\end{equation}

\subsection{ Simplification of the propagated state $| \Psi_t \rangle $ }
This is a quite cumbersome and complicated derivation. We sketch how
to obtain the final expression of this section. One writes the
propagator:
\begin{equation}
e^{ i Ht } = e^{ i S } e^{ i H_0 t }  e^{ -i S }
\end{equation}
Then one makes the effort to bring to the left the operator
$\psi^{\dag} (x)  $ using Eq.(\ref{LM}) also for obtaining the
expression of $ e^{ -i S } \psi^{\dag} e^{ i S } $ which is the same
result of Eq.(\ref{LM}) ( derived for $ e^{ i S } \psi^{\dag}_{+}
e^{ -i S } $) with $ \theta_q \to -\theta_q$, namely
$$h_q^{LM}(\theta_q)  \to \bar h_q^{LM}  = h_q^{LM} (-\theta_q). $$
Moreover we have to use that
\begin{eqnarray}
e^{ i H_0 t }  \psi^{\dag}_+ (x)  e^{-i H_0 t } & =& \int
dx^{\prime} \left( \sum_p  {1 \over L } e^{ i p (x-x^\prime) +i v_s
p t }
\right) \psi^{\dag}_+ ( x^\prime) \nonumber  \\
& = & \psi^{\dag}_+ ( x +v_s t)   \\
e^{ i H_0 t }  a^{\dag}_q   e^{-i H_0 t } &=& e^{i E_{q} t }
a^{\dag}_q   \\
\end{eqnarray}
and also the fact that
$$ e^{iS} e^{ i H_0 t }  a^{\dag}_q   e^{-i H_0 t } e^{-iS}
= e^{ i H t } e^{iS} a^{\dag}_q e^{-iS}  e^{-i H t } =  $$
$$= e^{i E_{q} t } cosh(\theta_q) a^{\dag}_q +
e^{-i E_{q} t } sinh(\theta_q) a_{-q} $$ where $E_q= v_s |q|$ is not
changed by the interaction. Then we arrive  to  the final result by
using  that  $e^{ i H_0 t } |FS\rangle = |FS \rangle$, repeated
applications of the Baker Haussdorf Campbell formula, and little
extra effort such as
$$ e^{ \alpha  a_q } e^{ \beta a^{\dag}_q } =
e^{ \beta a^{\dag}_q}  e^{ \alpha  a_q } e^{ \alpha \beta} $$ (a
relation that determines the constant $C_{1,2}(t)$ below):
\begin{eqnarray*}
\lefteqn{|\Psi(k,t) \rangle=} \\
&& M C_1  C_2(t) C_{12} (t)
 \int\limits_0^L dx ~e^{-i k x} \psi^{\dag}_+ (x + v_s t)
  \nonumber \\
 &\times& {\rm exp }\left[   \sum_{q} e^{i q x }
   \left( (h^{LM}_q+f_q h^{LM}_{-q}) e^{ i q v_s t } + B_q(t)  \right)
a^{\dag}_q \right] \nonumber  \\
&\times& e^{F}  |FS \rangle ~~~{\rm where: } \nonumber  \\
B_q(t) &=& \bar h_q^{LM} \left[  e^{i E_q t} cosh(\theta_q)
+ e^{-i E_q t}  f_q sinh(\theta_q) \right]   \nonumber \\
C_{1,2} (t) &=& {\rm exp } \left[ \sum_q h_{q}^{LM}  e^{ -i q v_s t}
B_q(t)\right]
\nonumber  \\
M &=& {\rm exp } \left[ 1/2 \sum_q \left( \bar h_{q}^{LM} \right)^2
\right]
\nonumber  \\
C_1 &=& {\rm exp }  \left[ 1/2 \sum_q h_q^{LM} (h_q^{LM}+h_{-q}^{LM}
f_q) \right]
\nonumber  \\
C_2 (t) &=& {\rm exp } \left[ 1/2 \sum_q
e^{-i E_q t} \bar h_{-q}^{LM} sinh(\theta_q) B_q(t)   \right] \nonumber\\
\end{eqnarray*}
where the constant $M$ follows from the normal order of
$$ {\rm exp } \left[
\sum_q e^{iqx} \left( \bar h_q^{LM}  a^{\dag}_q  + \bar h_{-q}^{LM}
a_{-q} \right) \right] $$,  the constant $C_1$ follows from the
normal ordering of
$$ e^{-F}
{\rm exp }\left[ \sum_q e^{iq (x+ v_s t) }
  ( h_q^{LM}  a^{\dag}_q  +h_{-q}^{LM} a_{-q} ) \right] e^F $$
and $C_2(t)$ from the normal ordering of:
$$ e^{-F} {\rm exp}
\left[ \sum_q  e^{i q x } \bar h_q^{LM}
 \left(e^{ i E_q t}  cosh(\theta_q) a^{\dag}_q  + \right. \right. $$
$$ \left. \left.
 + e^{ -i E_q t} sinh(\theta_q) a_{-q} \right)  \right] e^F $$
that can be made explicit by  using simple manipulations already
introduced in the previous section (see Eq.\ref{applyF}).

It is clear therefore that, exactly  as in the previous section,
 the above state can be
written as a generalized Jastrow Slater of the form $\Psi_{JhC}$,
with appropriate choice of the complex time dependent function $h_q(t)$ and
time dependent constant $C(t)$. Indeed after the simple replacement
in the dummy integration in $dx$ of $x+v_s t\to x$ (notice  that we
are using
 PBC and therefore $\psi^{\dag}_+(x+L)= \psi_+^\dag(x)$) so that the
integration domain can be shifted by arbitrary constants), we obtain
the following simple conditions to match:
\begin{eqnarray}
C(t) C_{Jh} (t) &=& e^{ i k v_s t } M C_1 C_2(t) C_{12}(t)  \\
{ i q  z_q^{(2)}(t)  (1+ f_q)
 \over \sqrt{ 2 \pi |q|} }
 & =&  e^{-i q v_s t}
 \left[ (h^{LM}_q+f_q h^{LM}_{-q}) e^{ i q v_s t } + B_q(t)\right] \nonumber \\
\end{eqnarray}
whereas the density density term in the Jastrow factor is always
characterized by  the same $v_q$ given by Eq.(\ref{vqdef}).



\end{document}